\documentclass[preprintnumbers,nofootinbib,superscriptaddress,10pt]{revtex4}
\usepackage[dvipdfmx]{graphicx} 
\usepackage{amsmath,amssymb,amsfonts, bm} 

\def\a{\alpha} \def\b{\beta} \def\g{\gamma}  \def\d{\delta}     \def\th{\theta}   \def\l{\lambda}        \def\r{\rho}          

  \def\nn{\nonumber}

\def\abs#1{\left| #1\right|}

\renewcommand{\Im}{{\rm Im}\,}

\newcommand{\Diag}[3]{ \begin{pmatrix} #1 & 0 & 0 \\ 0 & #2 & 0 \\ 0 & 0 & #3 \\\end{pmatrix}}

\usepackage{color}

\begin{document}

\title{\large 
Rephasing Invariant Formula for CP Phase in  Kobayashi--Maskawa Parametrization 
and Exact Sum Rule with Unitarity Triangle $\delta_{\rm PDG} + \delta_{\rm KM} = \pi - \alpha + \gamma$}

\preprint{STUPP-25-284}

\author{Masaki J. S. Yang}
\email{mjsyang@mail.saitama-u.ac.jp}
\affiliation{Department of Physics, Saitama University, 
Shimo-okubo, Sakura-ku, Saitama, 338-8570, Japan}
\affiliation{Department of Physics, Graduate School of Engineering Science,
Yokohama National University, Yokohama, 240-8501, Japan}



\begin{abstract} 

In this letter, we obtain a rephasing invariant formula for the CP phase in the Kobayashi--Maskawa parameterization $\delta_{\rm KM} =  \arg [ - { V_{ud} \det V_{\rm CKM} / V_{us} V_{ub} V_{cd} V_{td}} ]$. 
General perturbative expansion of the formula and observed value  $\delta_{\rm KM} \simeq \pi/2$
reveal that the phase difference of the 1-2 quark mixings 
$e^{i (\rho_{12}^{d} - \rho_{12}^{u})}$  is close to maximal for sufficiently small 1-3 quark mixings 
$s_{13}^{u,d}$. 
Moreover, combining this result with another formula for the CP phase $\delta_{\rm PDG}$ in the Particle Data Group parameterization, we derived an exact sum rule $\delta_{\rm PDG} + \delta_{\rm KM} = \pi -  \alpha + \gamma$ which relating the phases and the angles $\alpha, \beta, \gamma$ of the unitarity triangle. 

\end{abstract} 

\maketitle

\section{Introduction}

The Kobayashi--Maskawa (KM) theory \cite{Kobayashi:1973fv} is a framework for explaining CP violation in flavor mixing and plays a crucial role in the Standard Model. While the parametrization of the flavor mixing matrix is commonly expressed in the standard PDG parameterization by Chau and Keung  \cite{Chau:1984fp}, the original KM parametrization employed a different definition. 
In this KM parametrization, the CP-violating phase $\delta_{\rm KM}$ is known experimentally to be close to the  maximal value $\pi/2$ \cite{Koide:2004gj, Koide:2008yu}.
Relationships between this nearly maximal KM phase and the unitarity triangle have been a subject of discussion 
\cite{Hocker:2006xb, Frampton:2010ii, Dueck:2010fa, Frampton:2010uq,Li:2010ae,Qin:2011bq,Zhou:2011xm,Qin:2011ub,Qin:2010hn, Zhang:2012ys, Li:2012zxa, Zhang:2012bk}, in connection with the general treatment of the unitarity triangle \cite{Wu:1994di, Xing:2009eg, Harrison:2009bz, He:2013rba, He:2016dco, Xing:2019tsn, Harrison:2025rkp}. 
However, the phase difference between the KM and PDG parametrizations,
 and relations among the phases and angles of the unitarity triangle have not been clearly explained.

In this letter, we first express the CP phase $\delta_{\rm KM}$ by a rephasing invariant formula written by 
 arguments of matrix elements and the determinant of the mixing matrix $\det V$ \cite{Yang:2025hex,Yang:2025law}.  
Such invariants in which $\det V$ is fixed to unity have also been discussed in Refs.~\cite{Chang:2002yr,Kuo:2005pf,Chiu:2012uc,Chiu:2015ega,Chiu:2017ckv,Kuo:2019psm}. 
Comparing this expression with the standard phase $\delta_{\rm PDG}$ in the PDG parametrization and the angles $\a, \b, \g$ of the unitarity triangle, we derive an exact sum rule connecting these phases and angles.

\section{Rephasing Invariant Formula for CP Phase in KM Parametrization}

Following our previous approach \cite{Yang:2025hex,Yang:2025law}, we derive a rephasing invariant formula for the Kobayashi--Maskawa phase $\delta_{\rm KM}$ in an arbitrary phase convention.
The Cabibbo--Kobayashi--Maskawa (CKM) matrix in the original KM parametrization, denoted as 
$V_{\rm CKM}^{1}$ is given by
\begin{align}
V^{1}_{\rm CKM} 
& =
\begin{pmatrix}
1 & 0 & 0 \\
0 & c_{2} & -s_{2} \\
0 & s_{2} & c_{2}
\end{pmatrix}
\begin{pmatrix}
c_1 & -s_1 & 0 \\
s_1 & c_1 & 0 \\
0 & 0 & e^{i\delta_{\rm KM}}
\end{pmatrix} 
\begin{pmatrix}
1 & 0 & 0 \\
0 & c_{3} & s_{3} \\
0 & s_{3} & - c_{3}
\end{pmatrix} \nn \\
& = 
\begin{pmatrix}
c_1 & -s_1 c_3 & -s_1 s_3 \\
s_1 c_2 & c_1 c_2 c_3 - s_2 s_3 e^{i\delta_{\rm KM}} & c_1 c_2 s_3 + s_2 c_3 e^{i\delta_{\rm KM}} \\
s_1 s_2 & c_1 s_2 c_3 + c_2 s_3 e^{i\delta_{\rm KM}} & c_1 s_2 s_3 - c_2 c_3 e^{i\delta_{\rm KM}}
\end{pmatrix} . 
\end{align}
As in the PDG parametrization, phase redefinitions of five quark fields together with the sign freedom in $\delta_{\rm KM}$ allow all $c_{i}$ and $s_{i}$ to be taken as positive, $c_{i}, s_{i} > 0$. 
The following six conditions specify 
the phase structure in this parametrization; 
\begin{align}
\arg V_{ud}^{1} = \arg (- V_{us}^{1}) = \arg (- V_{ub}^{1}) = \arg V_{c d}^{1} = \arg V_{t d}^{1} = 0 \, , ~~ 
\arg  \det V_{\rm CKM}^{1} = \delta_{\rm KM} + \pi \, .
\end{align}

To convert a CKM matrix $V_{\rm CKM}$ in an arbitrary basis
into its KM parametrization $V_{\rm CKM}^{1}$, 
let us remove unphysical phases by a general rephasing transformation, 
\begin{align}
V_{\rm CKM}^{1} = \Psi_{L}^{\dagger} V_{\rm CKM} \Psi_{R} \, . 
\end{align}
Here, $\Psi_{L,R} = {\rm diag}(e^{i \gamma_{(L,R)1}} \, , e^{i \gamma_{(L,R)2}} \,, e^{i \gamma_{(L,R)3}})$ denotes diagonal phase matrices, where $\gamma_{(L,R)i}$ are phases to be determined.
The inverse transformation 
given by $V_{\rm CKM} = \Psi_{L} V_{\rm CKM}^{1} \Psi_{R}^{\dagger}$ is expressed as 
\begin{align}
\begin{pmatrix}
V_{ud} & V_{us} & V_{ub}  \\
V_{cd} & V_{cs} & V_{cb}  \\
V_{td} & V_{ts} & V_{tb}  \\
\end{pmatrix}
= 
\Diag{e^{i \g_{L1}}}{e^{i \g_{L2}}}{e^{i \g_{L3}}}
\begin{pmatrix}
|V_{ud}| & - |V_{us}| & - |V_{ub}|  \\
|V_{cd}| & V_{cs}^{1} & V_{cb}^{1}  \\
|V_{td}| & V_{ts}^{1} & V_{tb}^{1} \\
\end{pmatrix}
\Diag{e^{- i \g_{R1}}}{e^{- i \g_{R2}}}{e^{- i \g_{R3}}} .
\end{align}
From a relation $\arg \det V_{\rm CKM} = \delta_{\rm KM} + \pi + \sum_{i} (\gamma_{Li} - \gamma_{Ri})$ ,
we obtain the formula for the phase $\delta_{\rm KM}$ 
\begin{align}
\delta_{\rm KM} & = \arg  \det V_{\rm CKM} - \pi - 
 (\g_{L1} + \g_{L2} + \g_{L3} - \g_{R1} - \g_{R2} - \g_{R3}) \nn  \\
& = \arg  \det V_{\rm CKM} - 
 \arg (- V_{us} V_{ub} V_{cd} V_{td} / V_{ud}) 
= \arg \left[ - { V_{ud} \det V_{\rm CKM} \over  V_{us} V_{ub} V_{cd} V_{td}} \right] . 
\end{align}
This formula is manifestly rephasing invariant. Except for this particular combination, 
nontrivial phases of $V_{cs}^{1}$, $V_{cb}^{1}$, $V_{ts}^{1}$, $V_{tb}^{1}$ explicitly remain, and 
$\delta_{\rm KM}$ cannot be solved explicitly.
Remarkably, such a concise calculation had never been demonstrated 
in more than half a century, including the era of the $B$ factories.  

The parameters of the CKM matrix from the latest UTfit are given by \cite{UTfit:2022hsi} 
\begin{align}
\sin \th_{12} &= 0.22519 \pm 0.00083 \, ,  ~~~ \sin \th_{23} = 0.04200 \pm 0.00047 \, , \nn \\
\sin \th_{13} &= 0.003714 \pm 0.000092 \, ,  ~~~ \d = 1.137 \pm 0.022  = 65.15^{\circ} \pm1.3^{\circ}  \, . 
\end{align}
From the best fit values, the CP phase $\delta_{\rm KM}$ is evaluated from the standard PDG parametrization $V_{\rm CKM}^{0}$ as
\begin{align}
\d_{\rm KM } = \arg \left[ - { V_{ud}^{0} \det V_{\rm CKM}^{0} \over  V_{us}^{0} V_{ub}^{0} V_{cd}^{0} V_{td}^{0} } \right] = 87.56^{\circ} \, , ~~ \cos \d_{\rm KM} = 0.0425\, , ~~ \sin \d_{\rm KM} = 0.9991 \, . 
\label{obsdkm}
\end{align}
Here the principal value of $\arg$ lies in $[0, 2\pi)$. 
This agrees with the value obtained from moduli of matrix elements.
\begin{align}
\cos \d_{\rm KM} = {
|V_{ud}^{0} V_{us}^{0} V_{cd}^{0}|^{2} + |V_{ub}^{0} V_{td}^{0}|^{2} - |V_{cs}^{0}|^{2} ( 1 - |V_{ud}^{0}|^{2})^{2}
\over 2 |V_{ud}^{0} V_{us}^{0} V_{cd}^{0} V_{ub}^{0} V_{td}^{0} | } = 0.0425 \, . 
\end{align}
Furthermore, the relation with the Jarlskog invariant \cite{Jarlskog:1985ht} is given by 
\begin{align}
J = \Im [V_{ud} V_{cs} V_{us}^{*} V_{cd}^{*}] = 
c_{1} c_{2} c_{3} s_{1}^2 s_{2} s_{3} \sin \d_{\rm KM} \, . 
\end{align}
From this, $\sin \delta_{\rm KM}$ is evaluated as
\begin{align}
\sin \d_{\rm KM} = {J  \over c_{1} c_{2} c_{3} s_{1}^2 s_{2} s_{3} }
= {J (1- |V_{ud}|^{2}) \over \abs{V_{ud} V_{us} V_{cd} V_{ub} V_{td}}} 
= 0.9991 \, , 
\end{align}
and $\delta_{\rm KM}$ evaluated from the moduli of the matrix elements indeed lies in the first quadrant. 
Since the phase $\delta_{\rm KM}$ is nearly maximal $\pi/2$, it is intriguing to investigate its theoretical origin.

\subsection*{General Perturbative Expansion}

Here, we perform a general perturbative expansion of the KM phase to extract the physical interpretation.
The CKM matrix is defined as a misalignment between the diagonalizing matrices of the up-type $U_u$ and down-type quarks $U_d$.
These matrices are expressed as $U_{u,d} = \Phi_{u,d}^{L} U_{u,d}^{0} \Phi_{u,d}^{R}$ by phase matrices $\Phi_{u,d}^{L,R}$ and their PDG parametrizations $U_{u,d}^{0}$. 
By choosing an appropriate basis, each element of $U_{u,d}$ becomes of the same order as the corresponding CKM matrix element without loss of generality.
Therefore,  the mixing angles $s_{ij}^{u,d}$ of $U_{u,d}$ are  small parameters given by
\begin{align}
s_{12}^{u,d} \sim \l , ~~ s_{23}^{u,d} \sim \l^{2} , ~~  s_{13}^{u,d} \sim \l^{3} \, ,
\end{align}
with $\l = 0.2$. 
Moreover, since the unitary matrices can be redefined as
$U_{u,d}^{1} = \Phi_{u,d}^{L} U_{u,d}^{0} \Phi_{u,d}^{L \dagger} $
by the freedom of right-handed phase transformations, they can be expressed as follows:
\begin{align}
U_{u,d}^{1} = 
\begin{pmatrix}
1 & 0& 0 \\
0& c_{23}^{u,d} & s_{23}^{u,d}  e^{- i \r_{23}^{u,d} } \\
0& - s_{23}^{u,d}  e^{i \r_{23}^{u,d} } & c_{23}^{u,d}  \\
\end{pmatrix} 
\begin{pmatrix}
c_{13}^{u,d}  & 0 & s_{13}^{u,d}  e^{- i \r_{13}^{u,d} } \\
0 & 1 & 0 \\
- s_{13}^{u,d}  e^{i \r_{13}^{u,d} }& 0 & c_{13}^{u,d}  \\
\end{pmatrix} 
\begin{pmatrix}
c_{12}^{u,d}  & s_{12}^{u,d} e^{- i \r_{12}^{u,d} } & 0\\
- s_{12}^{u,d}  e^{i \r_{12}^{u,d} }& c_{12}^{u,d}  &0 \\
0 & 0 & 1 \\
\end{pmatrix}  ,
\end{align}
where $\r_{ij}^{u,d}$ are associated CP-violating phases to the mixing angles.
The leading order of the unitary matrices is approximated as 
\begin{align}
U_{u,d}^{1} 
 \simeq 
\begin{pmatrix}
1 & s_{12}^{u,d} e^{- i \r_{12}^{u,d} } & s_{13}^{u,d}  e^{- i \r_{13}^{u,d} } \\
- s_{12}^{u,d}  e^{i \r_{12}^{u,d} }& 1 & s_{23}^{u,d}  e^{- i \r_{23}^{u,d} } \\
- s_{13}^{u,d}  e^{i \r_{13}^{u,d} } + s_{12}^{u,d} s_{23}^{u,d}   e^{i \r_{12}^{u,d} + i \r_{23}^{u,d} }& - s_{23}^{u,d}  e^{i \r_{23}^{u,d} } & 1 \\
\end{pmatrix} .
\label{perturb}
\end{align}
The terms of the next-to-leading order are suppressed by at least $O(\lambda^{2})$  in each matrix element. Since the right-handed phases of the quarks $\Phi^{R}_{u,d}$ do not contribute to the CP phase, they are omitted hereafter.

In this case, the CKM matrix to be analyzed is redefined by $V_{\rm CKM} \equiv U_{u}^{1\dagger} U_{d}^{1}$.
Expanding each matrix element perturbatively in $\lambda$, we obtain
\begin{align}
\arg V_{ud} & \simeq  0 + O(\l^{2}) \, , \\
\arg V_{us} & \simeq \arg \left[  s^d_{12} e^{-i \rho^d_{12}} - s^u_{12} e^{-i \rho^u_{12}} \right]+ O(\l^{2}) \, ,   ~~~ \arg V_{cd}  \simeq  \arg (- V_{us}^{*}) + O(\l^{2}) \, , \\
\arg V_{ub} & \simeq  \arg \left[ s^d_{13} e^{-i \rho^d_{13}} - s^u_{13} e^{-i \rho^u_{13}}
- s^u_{12} e^{-i \rho^u_{12}} (s^d_{23} e^{-i \rho^d_{23}} - s^u_{23} e^{-i \rho^u_{23}} ) \right]+ O(\l^{2})\, , \\
 \arg V_{td} & \simeq  \arg \left[ - s^d_{13} e^{i \rho^d_{13}} + s^u_{13} e^{i \rho^u_{13}}
+ s^{d}_{12} e^{i \rho^{d}_{12}} (s^d_{23} e^{ i \rho^d_{23}} - s^u_{23} e^{i \rho^u_{23}} ) \right]+ O(\l^{2})\, .
\end{align}
Contributions of the CP phases $\r^{u,d}_{ij}$ appear at first order of $s_{ij}^{u,d}$ 
bacause these phases vanish in the limit $s_{ij}^{u,d} \to 0$.  
Note that the observed mixing angles $s_{ij}$ and $c_{ij}$ constrain the absolute values of the matrix elements as 
\begin{align}
|V_{us}| =  s_{12} c_{13} \, , ~~
|V_{ub}| = s_{13} \, , ~~
|V_{cb}| = s_{23} c_{13} \, .
\end{align}

With the present definition $\det V_{\rm CKM} = 1$, 
the rephasing invariant formula provides a general perturbative expression for the KM phase 
\begin{align}
\d_{\rm KM} &= 
\arg \left[ - { V_{ud} \det V_{\rm CKM} \over  V_{us} V_{ub} V_{cd} V_{td}} \right]
=
\arg \left[ + { V_{ub}^{*} / V_{td}} \right]+ O(\l^{2}) \, ,  \\
\abs {V_{ub} / V_{td}} e^{i \d_{\rm KM}} & =
- \frac{ s^d_{13} e^{i \rho^d_{13}} - s^u_{13} e^{i \rho^u_{13}}
- s^u_{12} e^{i \rho^u_{12}} (s^d_{23} e^{i \rho^d_{23}} - s^u_{23} e^{i \rho^u_{23}} ) } 
{ s^d_{13} e^{i \rho^d_{13}} - s^u_{13} e^{i \rho^u_{13}}
- s^{d}_{12} e^{i \rho^{d}_{12}} (s^d_{23} e^{ i \rho^d_{23}} - s^u_{23} e^{i \rho^u_{23}} ) }  + O(\l^{2})  \, .
\end{align}
This expression is sufficiently accurate because errors $O(\lambda^{2}) \simeq 4\%$ are comparable to the experimental uncertainty. 

In particular, the sum of the numerator and denominator is also constrained by the unitarity relation $V_{ud} V_{ub}^{*} + V_{cd} V_{cb}^{*} + V_{td} V_{tb}^{*} = 0$ as 
\begin{align}
V_{ub}^{*} + V_{td} & \simeq - V_{cd} V_{cb}^{*} \simeq  ( s^d_{12} e^{i \rho^d_{12}} - s^u_{12} e^{i \rho^u_{12}} ) (s^d_{23} e^{i \rho^d_{23}} - s^u_{23} e^{i \rho^u_{23}}) \, .
\end{align}
These conditions reduce to a problem in which differences among three complex numbers $a,b$, and $c$ are constrained; 
\begin{align}
|b - a| \simeq |V_{cd} V_{cb}^{*}| \, , ~ 
|c - b| \simeq |V_{td}| \, , ~ 
|a - c| \simeq |V_{ub}^{*}| \ , 
\end{align}
with $a = s^u_{12} e^{i \rho^u_{12}} V_{cb}^{*} , \, 
b = s^{d}_{12} e^{i \rho^{d}_{12}} V_{cb}^{*} , \, 
c = s^d_{13} e^{i \rho^d_{13}} - s^u_{13} e^{i \rho^u_{13}}$. 
Therefore, these numbers are represented by vectors on the complex plane from a certain point to the vertices of the unitarity triangle.

In such a definition, the almost maximal KM phase $\d_{\rm KM}$~(\ref{obsdkm}) essentially comes from the phase difference between $V_{td}$ and $V_{ub}$.
Many models predict $s^{d}_{12} \sim \sqrt{m_{d} / m_{s}} \sim 0.2$, and it is natural to consider that $s^{d}_{12} |V_{cb}| \sim 0.008$ dominates $V_{td}$. 
In this case, since the difference between the 1-3 mixings $s_{13}^{u,d}$ is required to be quite small, it can also be subdominant in $V_{ub}$.
Under such circumstances, the perturbative expression is simplified as 
\begin{align}
\abs {V_{ub} / V_{td}} e^{i \d_{\rm KM}} & \simeq  - \frac{s^{d}_{12} }{s^u_{12} } e^{i (\rho^u_{12} - \rho^{d}_{12} )} + O(\l^{2})  \, .
\end{align}
A rigorous treatment beyond such perturbation theory, especially concerning $s_{12}^{f}, s_{23}^{f}$, is also available in previous papers \cite{Yang:2024ulq, Yang:2025yst}. 

The implication of the almost maximal phase $e^{i (\rho^d_{12} - \rho^{u}_{12} )} \simeq -i$ has been discussed using various flavor symmetries and mass-matrix textures \cite{Shin:1985cg, Gronau:1985tx, Fritzsch:1985yv, Kang:1985nw,  Lehmann:1995br, Kang:1997uv, Fritzsch:1999im, Antusch:2009hq,Tanimoto:2015hqa, Xing:2003yj}. 
In particular, diagonal reflection symmetries \cite{Yang:2020qsa, Yang:2020goc, Yang:2021smh}, which are a specific class of generalized CP symmetries \cite{Ecker:1983hz, Gronau:1985sp, Feruglio:2012cw, Holthausen:2012dk}, always fix the phase difference of the 1-2 mixing to be $\pm \pi/2$.
\begin{align}
R \, m_{u,\nu}^{*} \, R = m_{u,\nu}, ~ m_{d,e}^{*} = m_{d,e} ~ {\rm with} ~ R = {\rm diag} (-1,1,1).
\end{align}
Since these GCPs fix the CP phases of the mass matrices as 
\begin{align}
m_{u} = 
\begin{pmatrix}
m_{u 11} & i \, m_{u 12} & i \, m_{u 13} \\
i \, m_{u 21} & m_{u 22} & m_{u 23} \\
i \, m_{u 31} & m_{u 32} & m_{u 33} \\
\end{pmatrix} , 
~~~
m_{d} = 
\begin{pmatrix}
m_{d 11} & m_{d 12} & m_{d 13} \\
m_{d 21} & m_{d 22} & m_{d 23} \\
m_{d 31} & m_{d 32} & m_{d 33} \\
\end{pmatrix} , 
\end{align}
with real parameters $m_{uij}$ and $m_{dij}$, 
the CKM matrix is generally constrained to the following form with the real rotation matrices $O_{u}$ and $O_{d}$ 
\begin{align}
V_{\rm CKM} = O_{u}^{T} \Diag{i}{1}{1} O_{d} \, .
\end{align}
Moreover, 
the sum rule that will be shown below implies that the experimental uncertainty in the KM phase $\d_{\rm KM}$ is expected to be $O(1^\circ)$ as that of phase $\d_{\rm PDG}$ and angles of the unitarity triangle $\alpha$, $\gamma$. Consequently, the observed value $\delta_{\rm KM} \simeq \pi/2$ is a robust constraint  that leaves little room for contributions from new physics.

\subsection*{Relationship between PDG Phase $\d_{\rm PDG}$ and Unitarity Triangle}

It is noteworthy to revisit relationships between the formula and the unitarity triangle.
From another invariant formula for the standard CP phase $\d_{\rm PDG}$ of PDG parameterization, 
\begin{align}
\d_{\rm PDG} = \arg \left[ \frac{ V_{ud} V_{us} V_{c b} V_{tb} }{ V_{ub} \det V_{\rm CKM} } \right] 
= 65.15^{\circ}  \, , 
\end{align}
the sum of the two phases
\begin{align}
\d_{\rm PDG} + \d_{\rm KM} &=  \arg \left[ \frac{ V_{ud} V_{us} V_{c b} V_{tb} }{ V_{ub} \det V_{\rm CKM} } \right]
+ 
 \arg \left[ - { V_{ud} \det V_{\rm CKM} \over  V_{us} V_{ub} V_{cd} V_{td}} \right] 
 =
  \arg \left[ - \frac{ V_{ud}^{2}  V_{c b} V_{tb} }{ V_{ub}^{2} V_{cd} V_{td}  } \right] \, ,  
\end{align}
leads to a new rephasing invariant.

This invariant is expressed by the angles $\a, \b$ and $\g$ of the unitarity triangle, 
\begin{align}
 \arg \left[ - \frac{ V_{ud}^{2}  V_{c b} V_{tb} }{ V_{ub}^{2} V_{cd} V_{td}^{}  } \right] 
 = 
 \arg \left [  { V_{ud}^{} V_{ub}^{*} \over V_{td}^{} V_{tb}^{*} } \right ] + 
 \arg \left[ -  \frac{ V_{ud}  V_{c b}  }{ V_{ub} V_{cd}  } \right] 
 =  \pi -  \a + \g =  2 \gamma + \b \, ,
\end{align}
where 
\begin{align}
\a = \arg \left [ - { V_{td}^{} V_{tb}^{*} \over V_{ud}^{} V_{ub}^{*} } \right ] = 92.40^{\circ}, ~~
\b = \arg \left [ - { V_{cd}^{} V_{cb}^{*} \over V_{td}^{} V_{tb}^{*} }  \right ] = 22.49^{\circ}, ~~
\g = \arg \left [ - { V_{ud}^{} V_{ub}^{*} \over V_{cd}^{} V_{cb}^{*} }  \right ] = 65.11 ^{\circ}.
\end{align}
This sum rule is an exact equality as long as $V_{\rm CKM}$ is unitary.
It is also confirmed numerically as 
\begin{align}
 \d_{\rm PDG} + \d_{\rm KM}
=  65.15^{\circ} + 87.56^{\circ} = 180^{\circ} - 92.40^{\circ} + 65.11 ^{\circ} = \pi -  \a + \g  \, . 
\end{align}
Although this $\pi$ appears superfluous,  
it originates from the determinant of the KM parametrization. 
By adopting a different parametrization, it is reduced to a simple phase difference.

Alternatively, in an equivalent form,
\begin{align}
 \d_{\rm PDG} - \g &= 
 \arg \left[ \frac{ V_{ud} V_{us} V_{c b} V_{tb} }{ V_{ub} \det V_{\rm CKM} } \right]
 - \arg \left [ - { V_{ud}^{} V_{ub}^{*} \over V_{cd}^{} V_{cb}^{*} }  \right ] 
 =  \arg \left[ -  { V_{us}  V_{cd} V_{tb} \over \det V_{\rm CKM} } \right] 
 =  0.035^{\circ}   \, , \nn \\
&=  \arg \left[ { V_{us} V_{ub} V_{cd} V_{td} \over V_{ud} \det V_{\rm CKM} } \right] - \arg \left [ - { V_{td}^{} V_{tb}^{*} \over V_{ud}^{} V_{ub}^{*} } \right ]
= \pi - \d_{\rm KM} - \a \, .   
\end{align}
As discussed in our previous work \cite{Yang:2025law},  
it leads to an equality between another invariants and the differences of phases and angles.
Since the argument $\arg \left[ -  { V_{us}  V_{cd} V_{tb} / \det V_{\rm CKM} } \right] $ is a small quantity of $O(\l^{4})$, an implication of this sum rule is 
\begin{align}
 \d_{\rm PDG}  \fallingdotseq \g \, , ~~~
 \d_{\rm KM} + \a \fallingdotseq  \pi \, .
\end{align}
Although they are well-known relations, the fact that their differences are expressed in a simple form of ``third-order'' invariants \cite{Chang:2002yr,Kuo:2005pf} will provide new insight into studies of CP violation.

\section{Summary}

In this letter, by removing unphysical phases of the CKM matrix $V_{\rm CKM}$, 
we obtain a rephasing invariant formula for the CP phase
$\delta_{\rm KM} =  \arg [ - { V_{ud} \det V_{\rm CKM} / V_{us} V_{ub} V_{cd} V_{td}} ]$  in the Kobayashi--Maskawa parameterization. 
General perturbative expansion of the formula and observed value $\delta_{\rm KM} \simeq \pi/2$
reveal that the phase difference of the 1-2 quark mixings 
$e^{i (\rho_{12}^{d} - \rho_{12}^{u})}$ is close to maximal for sufficiently small 1-3 quark mixings 
$s_{13}^{u,d}$. 
Moreover, combining this result with another formula for the CP phase $\delta_{\rm PDG}$ in the PDG parameterization, we derived an exact sum rule $\d_{\rm PDG} + \d_{\rm KM} = \pi -  \a + \g$ which relating the phases and the angles $\a, \b, \g$ of the unitarity triangle. 

These results provide a new understanding of relations between different parameterizations of the flavor mixing  matrix, as well as a theoretical perspective for discussing the unitarity triangles and CP phases. 
Similar equalities and sum rules are expected to hold for other parameterizations of the mixing matrix and other unitarity triangles as well.

\section*{Acknowledgment}

The study is partly supported by the MEXT Leading Initiative for Excellent Young Researchers Grant Number JP2023L0013.


\end{document}